\documentclass{webofc}
\usepackage[varg]{txfonts}   
\usepackage{tabularx,booktabs}
\usepackage{sidecap}
\usepackage{lineno}
\usepackage{hyperref}
\usepackage{xspace}
\usepackage{color}
\usepackage{xcolor}
\usepackage{rotating}
\definecolor{light-gray}{gray}{0.8}

\newcommand{\cmnt}[1]{}

\newcommand{\pt}{\ensuremath{p_{\rm T}}\xspace}

\newcommand{\sqrtsNN}{\ensuremath{\sqrt{s_{\rm NN}}}\xspace}

\newcommand{\dnchdetacubic}{\ensuremath{\average{{\rm d}N_{\rm ch}/{\rm d}\eta}^{1/3}}\xspace}

\newcommand{\yless}[1]{\ensuremath{\left|y\right| < #1}\xspace}

\newcommand{\GeVc}{GeV/$c$\xspace}

\newcommand{\pp}{\ensuremath{\rm pp}\xspace}
\newcommand{\pPb}{p--Pb\xspace}
\newcommand{\PbPb}{Pb--Pb\xspace}

\newcommand{\average}[1]{\ensuremath{\langle #1 \rangle}\xspace}

\newcommand{\Lambdastar}{\ensuremath{\Lambda(1520)}\xspace}

\newcommand{\Kstar}{\ensuremath{{\rm K}^{\ast}(892)^{0}}\xspace}

\newcommand{\rhomeson}{\ensuremath{\rho(770)^{0}}\xspace}
\newcommand{\Xistar}{\ensuremath{\Xi^{\ast 0}}\xspace}

\begin{document}
\title{Probing the hadronic phase with resonances of different lifetimes in Pb-Pb collisions with ALICE}
\author{\firstname{Neelima} \lastname{Agrawal}\inst{1}\fnsep\thanks{\email{neelima.agrawal@cern.ch}} (on behalf of ALICE collaboration)}
\institute{Indian Institute of Technology Bombay, India}

\abstract{The ALICE experiment has measured the production of a rich set of hadronic resonances, such as \rhomeson, \Kstar, $\phi$(1020), $\Sigma^{\pm}$(1385), \Lambdastar and \Xistar in \pp, \pPb and \PbPb collisions at various energies at the LHC. A comprehensive overview and the latest results are presented in this paper. Special focus is given to the role of hadronic resonances for the study of final-state effects in high-energy collisions. In particular, the measurement of resonance production in heavy-ion collisions has the capability to provide insight into the existence of a prolonged hadronic phase after hadronisation. The observation of the suppression of the production of \Lambdastar resonance in central \PbPb collisions at \sqrtsNN = 2.76 TeV adds further support to the existence of such a dense hadronic phase, as already evidenced by the ratios \Kstar/K and \rhomeson/$\pi$.}
\maketitle
\section{Introduction}
\begin{figure}[t]
\centering
\sidecaption
\includegraphics[height=0.45\linewidth]{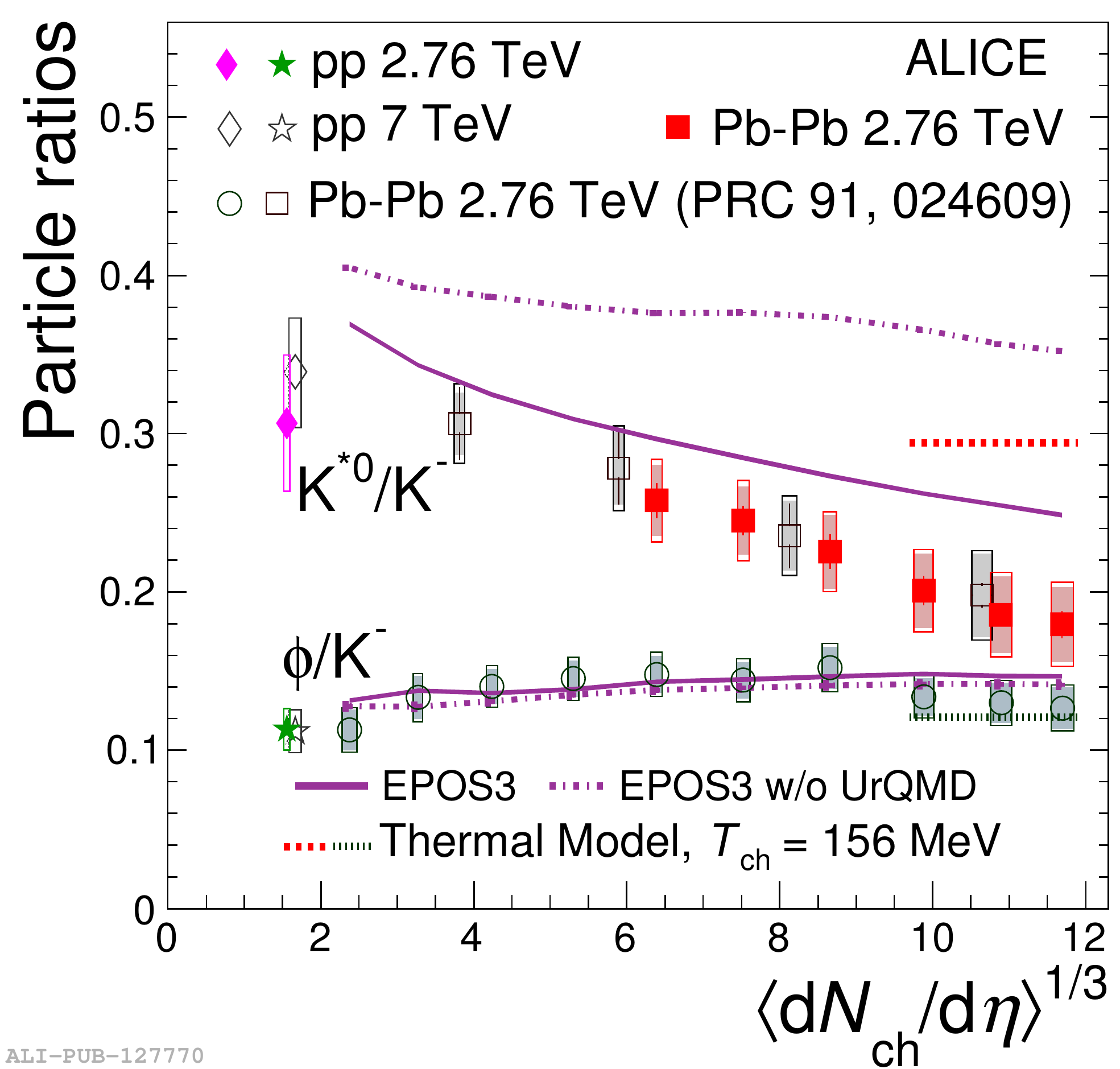}
\caption{The \pt-integrated ratios of \Kstar/K$^{-}$ and $\phi$(1020)/K$^{-}$ as a function of \dnchdetacubic measured at mid-rapidity in pp collisions at $\sqrt{s}$ = 2.76 TeV and 7 TeV and \PbPb collisions at \sqrtsNN = 2.76 TeV~\cite{Abelev:2012hy}. The statistical, total systematics and centrality uncorrelated systematic uncertainties are represented by the bars, empty boxes and shaded boxes, respectively. The expectations from a SHM calculation with a chemical freeze-out temperature of 156 MeV for the most central collisions and EPOS3 calculations are shown.}
\label{fig1}   
\vspace{-0.5cm}   
\end{figure}

Lattice QCD calculations predict that at high enough temperature ($T_{\rm c}\sim$156 MeV) and energy density (> 1 GeV/fm$^{3}$), nuclear matter undergoes a phase transition to a deconfined state of quarks and gluons known as the Quark-Gluon Plasma (QGP)~\cite{Karsch:2001cy}. These critical conditions are experimentally achieved by colliding heavy ions at ultra-relativistic energies~\cite{Roland:2014jsa,Ohnishi:2011aa}. In such collisions, a fireball of very hot and dense partonic matter is formed and persists for a very short time. The QGP matter cools down and transitions back to normal hadronic matter, which subsequently undergoes chemical and kinetic freeze-out. Due to their lifetimes shorter or comparable to the lifetime of the fireball ($\sim$ 10 fm/$c$~\cite{Aamodt:2011mr}), hadronic resonance are good probes to investigate the
properties of the hadronic medium existing in between the chemical and kinetic freeze-out. Resonances are produced at hadronisation and can decay inside the hadronic medium. If the hadronic phase lasts long enough, the decay daughters of very short-lived resonances experience its full evolution and suffer rescattering via elastic collisions in the dense hadronic medium, which could modify their correlations and hence the experimentally measured resonance yields. As a consequence, the reconstructible resonance
yield is lower than the original yield. Alternatively, depending upon the density
and duration of the hadronic medium, hadrons might interact
pseudoelastically to form resonances, which can increase the resonance yield.  The competing effects of
rescattering and regeneration on the final reconstructible resonance
yields is dependent on the time scale between
chemical and kinetic freeze-out, the relevant hadronic interaction cross sections and the resonance lifetime. The Statistical Hadronisation Model (SHM)
predicts ratios of resonance yields to stable particle yields at the chemical
freeze-out stage~\cite{Andronic:2005yp,Wheaton:2004vg,Petran:2013dva}. Deviation from the model predictions can be used to estimate the lifetime of the hadronic phase. Figure~\ref{fig1} shows the ratio of \Kstar ($\tau$ = 4.2 fm/$c$) and $\phi$ ($\tau$ = 46.2 fm/$c$) to kaons as a function of \dnchdetacubic~\cite{Abelev:2012hy}. The production of the short-lived \Kstar is suppressed in central \PbPb collisions, suggesting that the phenomenon might be due to the rescattering of the \Kstar decay products within the dense hadronic medium before the kinetic freeze-out. No suppression is observed for the $\phi$(1020) mesons which could indicate that it mostly decays outside the fireball due to its longer lifetime. 

\begin{table}[b]
  \caption{Hadronic resonances with their lifetime, decay width, valance quark content, decay mode exploited for the measurements and corresponding branching ratio (BR)~\cite{Olive:2016xmw}.}
  \label{tab1}
  \centering
  \begin{tabular*}{\textwidth}{@{\extracolsep{\fill}}lcccccc}
    \toprule
    resonances & $\tau$ (fm/$c$) & $\Gamma$ (MeV) & valance quark contents & decay mode & BR[\%]\\
    \midrule
    $\rho(770)^{0}$ & 1.3 & 149.1 $\pm$ 0.8 & $(u\overline{u} + d\overline{d})/ \sqrt{2}$ & $\pi ^{+} + \pi ^{-}$ & 100 \\
    \Kstar & 4.2 & 47.3 $\pm$ 0.5 & $d\overline{s}$ & K$^{+} + \pi ^{-}$ & 66.6 \\
    $\phi$(1020) & 46 & 4.3 $\pm$ 0.02 & $s\overline{s}$ & K$^{+}$ + K$^{-}$ & 48.9 \\ 
    $\Sigma$(1385)$^{+}$ & 5.5 & 36.0 $\pm$ 0.7 & $uus$ & p + $\pi^{-}$ + $\pi^{+}$ & 87.0 \\ 
    \Lambdastar  & 12.6 & 15.6 $\pm$ 1.0 & $uds$ & K$^{-}$ + p & 22.5 \\
    $\Xi(1530)^{0}$ & 21.7 & 9.1 $\pm$ 0.5 & $uss$ & p + 2$\pi^{-}$ + $\pi^{+}$ & 66.7 \\
    \bottomrule
  \end{tabular*}
\end{table}

\section{Analysis Method and Results}
\begin{figure}[t]
\centering
\includegraphics[height=0.36\linewidth,trim=0 -0.65cm 0 0.6cm]{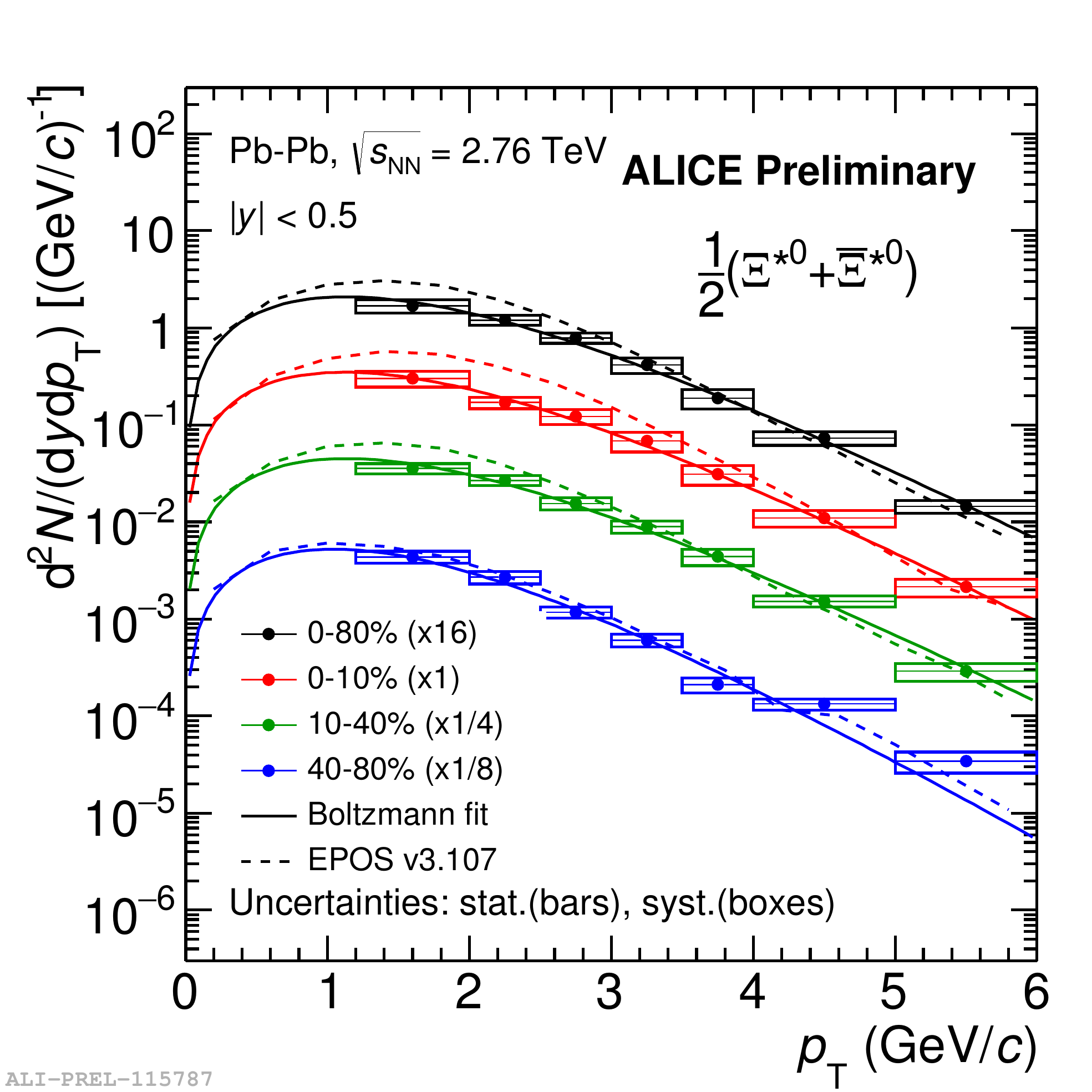}
\includegraphics[height=0.36\linewidth]{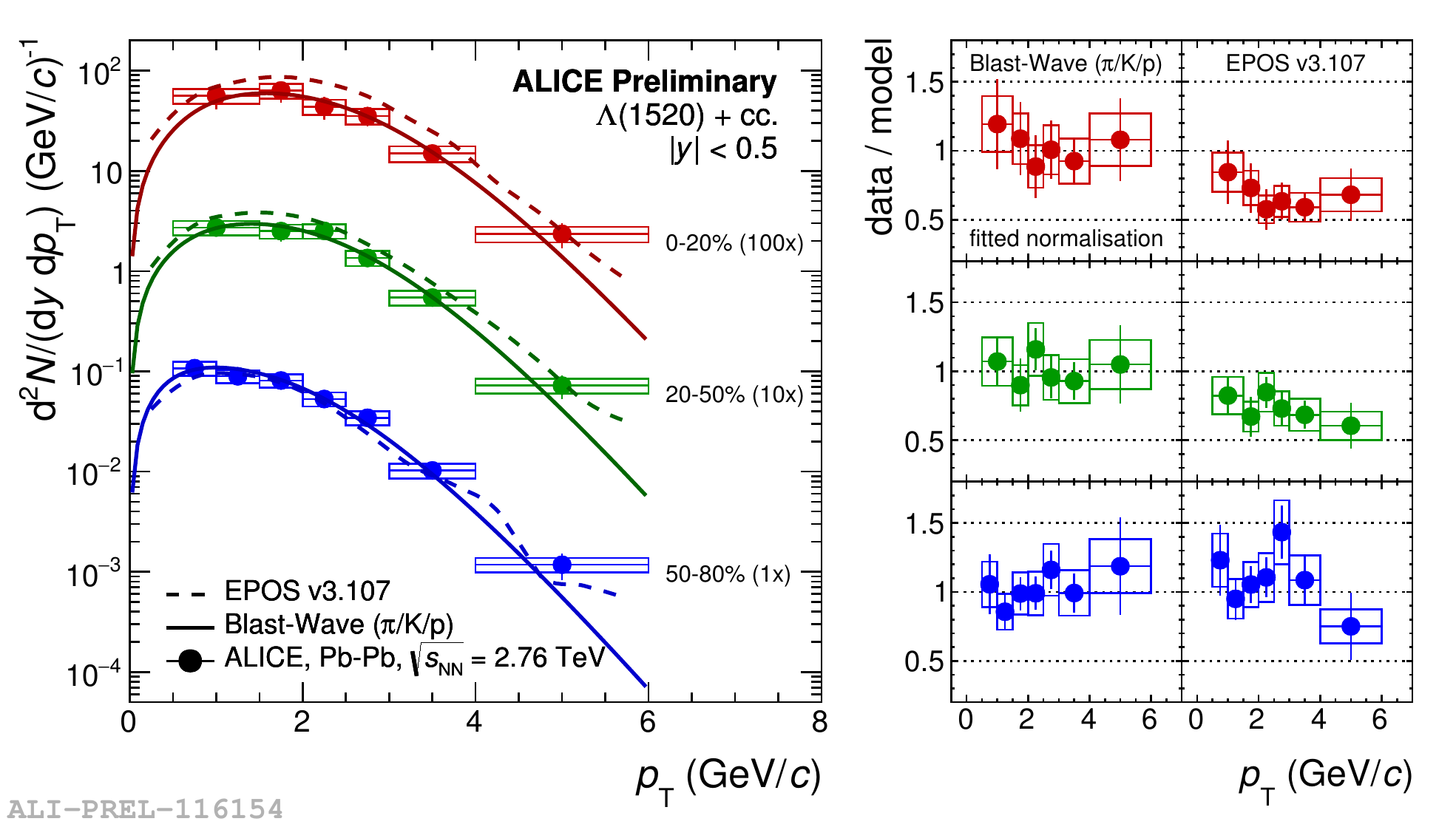}
\caption{The \pt spectra of \Xistar (left) and \Lambdastar (middle) in \PbPb collisions at \sqrtsNN = 2.76 TeV. The \Lambdastar spectra ratios with respect to Blast-Wave and EPOS3 predictions (right).}
\label{fig2}      
\end{figure}

Several hadronic resonances with different lifetimes have been measured by ALICE at midrapidity (\yless{0.5}) in \pp, \pPb and \PbPb collisions at various centre-of-mass energies. The properties of these resonances are listed in Table~\ref{tab1}. The resonance signal is reconstructed using an invariant mass technique. The decay products are identified using the information from the TPC and TOF detectors. The invariant mass distribution of the correct-sign pair of decay daughters is constructed for each centrality/multiplicity class and \pt interval. The combinatorial background is estimated from either mixed events or like-charged tracks from the same event and subtracted from the signal distribution; the result is then fitted to extract the resonance yield. The \pt-integrated yields are obtained for all resonances and collision systems and are corrected for the decay channel branching ratio, detector acceptance and reconstruction efficiency. For example, Figure~\ref{fig2} shows the spectral shapes of  \Xistar (left) ($\tau$ = 21.7 fm/$c$) and \Lambdastar (middle) ($\tau$ = 12.6 fm/$c$) in \PbPb collisions measured in the \pt range 1.2-6.0 \GeVc and 0.5-6.0 \GeVc, respectively. The spectra are fitted with appropriate parametrisations to obtain the particle yield in the unmeasured \pt region. The spectral shapes have also been compared with predictions from the EPOS3 model~\cite{Knospe:2015nva}, which describes the full evolution of heavy ion collisions and embeds the UrQMD model for the description of the hadronic phase. This model reproduces the spectral shapes reasonably well but over-predicts the yield for central \PbPb collisions. 

\begin{figure}[t]
\centering
\includegraphics[height=0.48\linewidth]{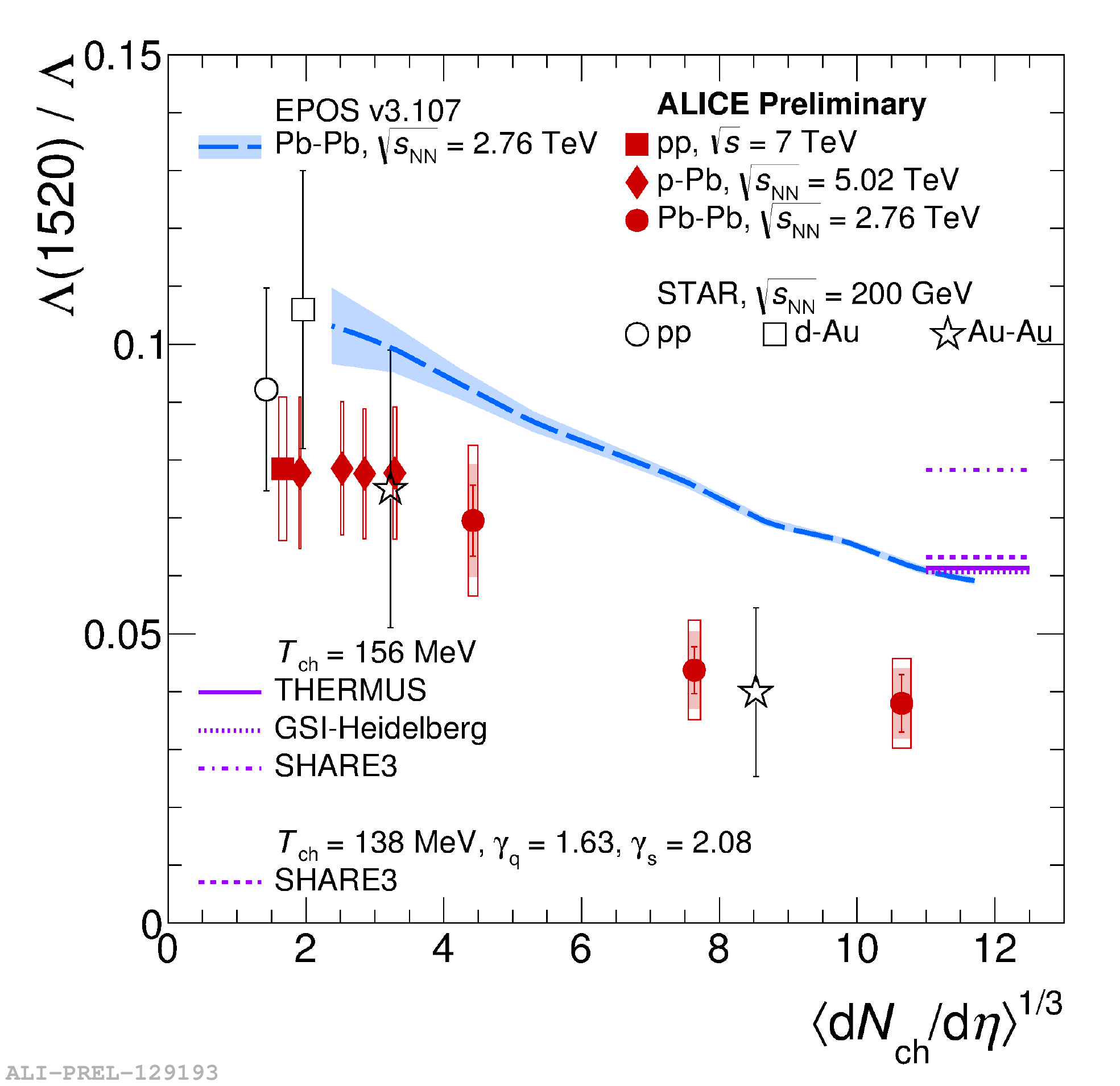}
\includegraphics[height=0.48\linewidth,trim=0 0.4cm 0 0]{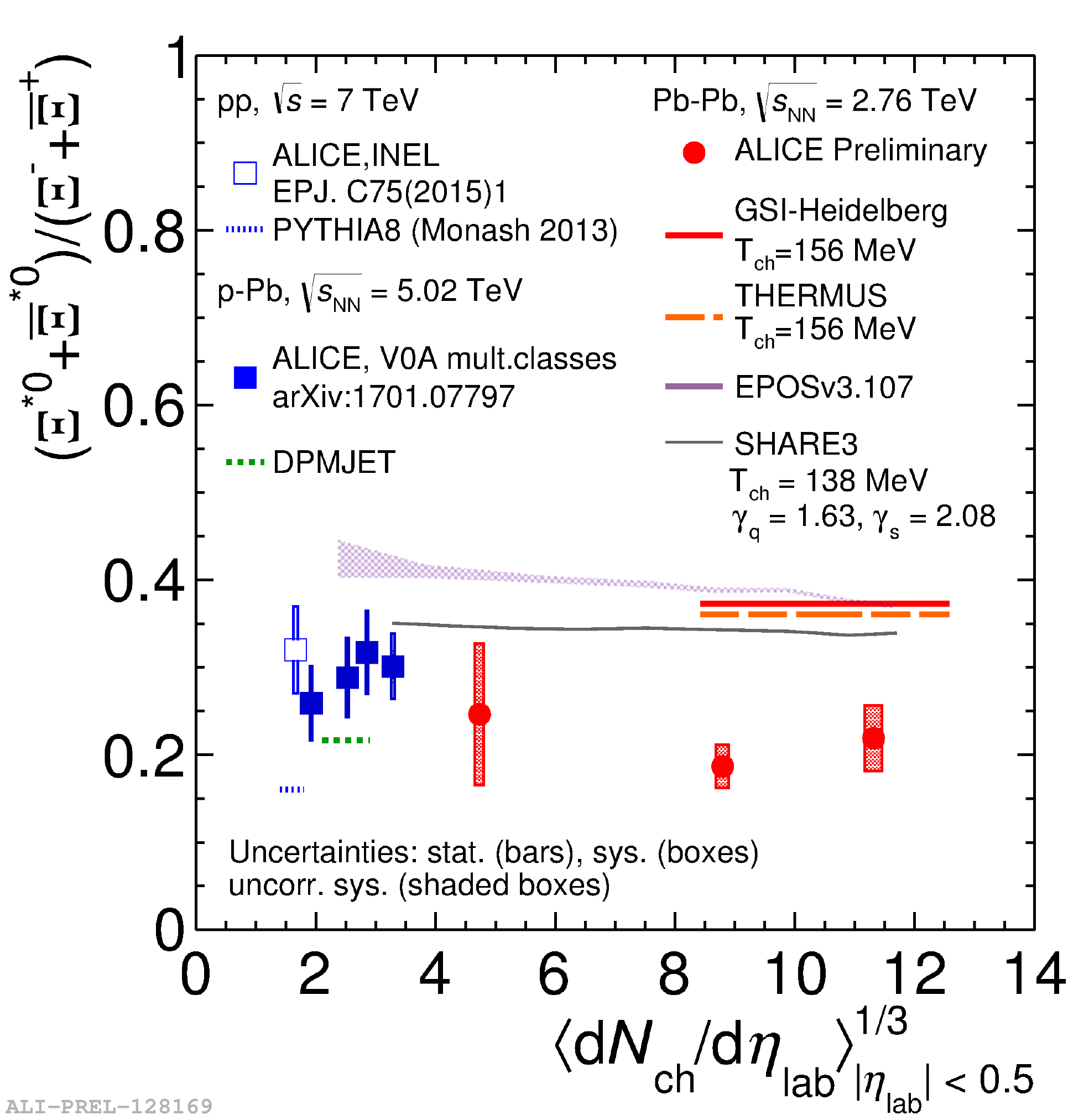}
\caption{The \pt-integrated ratio of \Lambdastar/$\Lambda$ (left) and \Xistar/$\Xi$ (right) as a function of \dnchdetacubic measured in \PbPb collisions at \sqrtsNN = 2.76. The statistical errors, systematic uncertainties and centrality uncorrelated systematic uncertainties are represented by the bars, empty boxes and shaded boxes, respectively. Predictions from several SHM models~\cite{Andronic:2005yp,Wheaton:2004vg,Petran:2013dva} as well as the prediction from EPOS3~\cite{Knospe:2015nva} are compared to the data.}
\label{fig3}      
\end{figure}

The integrated yield ratios of resonances to stable hyperons with the same strangeness content is calculated in order to investigate the impact of the hadronic phase on the reconstructed resonance yield. Figure~\ref{fig3} shows the \Lambdastar/$\Lambda$ (left) and \Xistar/$\Xi$ (right) ratios as a function of \dnchdetacubic in \pp, \pPb and \PbPb collisions at 7, 5.02 and 2.76 TeV, respectively. Predictions from Monte Carlo generators commonly used in \pp and \pPb collisions, EPOS3 and various SHM predictions~\cite{Andronic:2005yp,Wheaton:2004vg,Petran:2013dva} for \PbPb collisions are also shown. The \Lambdastar/$\Lambda$ shows suppression in central \PbPb collisions with respect to peripheral \PbPb and \pPb and \pp collisions. The SHM predictions for \PbPb collisions overestimate the data. On the other hand, EPOS3 qualitatively reproduces the suppression trend of ratio, however, it overestimates the value. The \Lambdastar/$\Lambda$ ratio suppression trend indicates the dominance of rescattering of the decay daughters over regeneration. The \Xistar/$\Xi$ ratio shows relatively weak suppression from \pp, \pPb, peripheral \PbPb to central \PbPb collisions. The ratio values are underestimated by PYTHIA8 in \pp, DPMJET in \pPb and overestimate by various SHM calculations in \PbPb collisions. In addition, EPOS3 with UrQMD predicts a slight decrease of the \Xistar/$\Xi$ ratio from peripheral to central Pb-Pb collisions and a yield value which is consistent with the SHM predictions for most central Pb-Pb. The trend is qualitatively consistent with the observed one, given the present uncertainties of the measurement, however the measured yield ratio is lower than the predictions.     

\section{Conclusion}
In this paper, we have presented the first measurement of \Lambdastar and \Xistar in \pp, \pPb and \PbPb collisions at $\sqrt{s}$ = 7, \sqrtsNN = 5.02 and 2.76 TeV, respectively. The \Lambdastar/$\Lambda$ ratio is suppressed in central collisions compared to the values observed in peripheral \PbPb, \pp and \pPb collisions. This is similar to the behaviour observed in the $\rho/\pi$ and \Kstar/K$^{-}$ ratios. The suppression of \Lambdastar in central \PbPb collisions suggests the dominance of (pseudo)elastic rescattering of decay daughter particles in the hadronic phase. Moreover, the suppressed value of \Lambdastar/$\Lambda$ ratio at highest \dnchdetacubic could be attributed to a larger fireball size and its life-time accessible at LHC energies. The \Xistar/$\Xi$ ratio exhibits a very weak suppression from \pp, \pPb, peripheral \PbPb to central \PbPb collisions. The \Lambdastar/$\Lambda$ and \Xistar/$\Xi$ ratios in central \PbPb collisions are compared and found to be lower than the calculations from several implementations of the Statistical Hadronisation Models in Grand Canonical description and one non-equilibrium model. The deviations of these ratios in the data as compared to the SHM models are useful to decode effects related to the interactions in hadronic medium and also to estimate the time span between chemical and kinetic freeze-out. These results further support the existence of a hadronic phase lasting long enough to cause a significant reduction of the reconstructible yield of short-lived resonances. 

\nocite{*}
\bibliography{SQMresonance.bib}
\end{document}